\documentclass{article}

\usepackage[utf8]{inputenc}

\usepackage{amsmath}
\usepackage{amssymb}
\usepackage{graphicx}
\usepackage{subcaption}
\usepackage{dsfont}  
\usepackage{authblk}
\usepackage{hyperref}
\usepackage{siunitx}
\usepackage{xcolor}

\usepackage{float}
\usepackage{booktabs}
\usepackage{adjustbox}
\usepackage{tikz}
\usepackage{csquotes}
\usepackage{multirow}
\usepackage[square,numbers,sort&compress]{natbib}

\usetikzlibrary{shapes, arrows, positioning,decorations.markings}
\tikzstyle{vecArrow} = [thick, decoration={markings,mark=at position
	1 with {\arrow[semithick]{open triangle 60}}},
double distance=1.4pt, shorten >= 5.5pt,
preaction = {decorate},
postaction = {draw,line width=1.4pt, white,shorten >= 4.5pt}]
\tikzstyle{innerWhite} = [semithick, white,line width=1.4pt, shorten >= 4.5pt]
\hypersetup{hidelinks}
\usepackage[tmargin=1in,bmargin=1in,lmargin=0.75in,rmargin=0.75in]{geometry}
\newcommand{\R}{\ensuremath{\mathbb{R}}}

{
	{

	}
}

\newcommand\blfootnote[1]{%
	\begingroup
	\renewcommand\thefootnote{}\footnote{#1}%
	\addtocounter{footnote}{-1}%
	\endgroup
}

\newcommand{\noise}{\sigma_\mathrm{noise}}
\newcommand{\mPred}{{M_\text{pred}}}
\newcommand{\aPred}{{\alpha_\text{pred}}}
\newcommand{\mPredPrime}{{M_\text{pred}^\prime}}
\newcommand{\aPredPrime}{{\alpha_\text{pred}^\prime}}
\newcommand{\Rtwo}{\text{R}^2}
\newcommand{\Lone}{\text{L}_1}

\DeclareMathOperator*{\argmin}{arg\,min}

\title{Statistical learning of structure-property relationships for transport in porous media, using hybrid AI modeling}
\author{Somayeh Hosseinhashemi$^{a,\ast}$, Philipp Rieder$^{b}$, Orkun Furat$^{b}$, Benedikt Prifling$^{b}$, Changlin Wu $^{a}$, Christoph Thon$^{a}$,  Volker Schmidt$^{b}$, Carsten Schilde$^{a}$}

\begin{document}
\setlength{\parindent}{0pt}
\maketitle
\vspace{-2em}
\begin{center}
	\it
	$^a$Institute of Particle Technology, Technical University of Braunschweig, Franz-Liszt Stra{\upshape{\ss}}e 35A, 38104~Braunschweig, Germany
	\\
	$^b$Institute of Stochastics, Ulm University, Helmholtzstra{\upshape{\ss}}e~18, 89069~Ulm, Germany
\end{center}
\blfootnote{$^\ast$Corresponding author: Somayeh Hosseinhashemi; Email: \textit{s.hosseinhashemi@tu-braunschweig.de}}

\begin{abstract}
\noindent
The 3D microstructure of porous media, such as electrodes in lithium-ion batteries and solid-oxide fuel cells or fiber-based materials, significantly impacts the resulting macroscopic properties, including effective diffusivity or permeability. Consequently, quantitative structure-property relationships, which link structural descriptors of 3D microstructures such as porosity, specific surface area, or geodesic tortuosity to effective transport properties, are crucial for further optimizing the performance of porous media. To overcome the limitations of 3D imaging, which is time-consuming and costly, parametric stochastic 3D microstructure modeling, which uses tools from stochastic geometry, is a powerful tool to generate many virtual but realistic structures at the cost of computer simulations. The present paper uses 90,000 virtually generated 3D microstructures, consisting of a solid phase and the pore space, derived from literature by systematically varying parameters of stochastic 3D microstructure models. Previously, this data set has been used to establish quantitative microstructure-property relationships utilizing analytical regression equations, artificial neural networks, and convolution neural networks. The present paper extends these findings by applying a hybrid AI framework to this data set. More precisely, symbolic regression, powered by deep neural networks, genetic algorithms, and graph attention networks, is used to derive precise and robust analytical equations. These equations model the relationships between structural descriptors and effective transport properties without requiring manual specification of the underlying functional relationship. By integrating AI with traditional computational methods, the hybrid AI framework not only generates predictive equations but also enhances conventional modeling approaches by capturing relationships influenced by specific microstructural features traditionally underrepresented. Thus, this paper significantly advances the predictive modeling capabilities in materials science and process engineering, offering vital insights for designing and optimizing new materials with tailored transport properties. 
\end{abstract}


\section{Introduction}

The 3D microstructure of various porous functional materials such as solar cells and electrodes in lithium-ion batteries or fuel cells significantly influences the resulting macroscopic properties such as effective diffusivity and permeability. This relationship is essential for optimizing the performance of porous media \cite{Torquato.2002, willot.2018}. Thus, a quantitative understanding of the interplay between the complex 3D microstructure, characterized via microstructural descriptors such as porosity, specific surface area or constrictivity, and effective macroscopic properties is required for further progress in various research fields, including battery technology and pharmaceutical systems \cite{MASARO1999731, Berryman1985}. Historically, traditional methods for establishing these relationships have involved extensive experimental efforts, often supported by high-resolution 3D imaging techniques like X-ray tomography and focused ion beam scanning electron microscopy. However, these methods are limited by high costs and time demands, prompting a shift towards virtual materials testing \cite{Dunn2020, Saunders2021}. This approach utilizes parametric stochastic 3D microstructure modeling, employing tools from stochastic geometry, to generate a vast dataset of virtual but realistic 3D microstructures at a fraction of the cost and time of experimental methods \cite{lantuejoul.2013, chiu.2013}.

\smallskip
\noindent
In this context, this study employs a comprehensive dataset of 90,000 virtual microstructures derived from systematic variations in the parameters of nine different types of stochastic 3D microstructure models \cite{prifling2021}. This dataset has facilitated a deeper understanding of microstructure-property relationships through traditional analytical methods and the emerging use of computational techniques such as artificial neural networks (ANNs) and convolutional neural networks (CNNs) \cite{prifling2021}. These advanced methods, including high-dimensional regression and machine learning (ML), have captured the complex effects of microstructures on material properties, marking significant advances in the automation and refinement of prediction processes \cite{wu.2019, WANG2020100035}. Despite these technological advances, challenges remain, especially in applying computational methods to large, heterogeneous datasets essential for robust generalizations across different material types and conditions. To bridge this gap, this study introduces a hybrid AI framework that combines the strengths of symbolic regression (SR), powered by deep neural networks (DNNs), genetic algorithms (GAs), and graph attention networks (GATs) to derive robust and accurate analytical microstructure-property relationships. This framework not only adapts to but also learns from the complexity of microstructural data, thus providing transformative insights into the design and optimization of materials \cite{HOSSEINHASHEMI2023,hosseinhashemi2024}. Specifically, this framework incorporates a novel dual implementation of GATs within the SR process. The first implementation uses GATs to focus on data augmentation through graph-based learning to enrich the training dataset. The second implementation uses GATs to analyze feature interactions, where "features" refer to the microstructural descriptors that are critical to understanding effective material properties. By integrating these GAT-based approaches with SR techniques, our framework can both expand the available training data and capture complex interactions between microstructural descriptors, leading to more accurate and physically interpretable microstructure-property relationships.

\smallskip
\noindent
Our approach refines prediction equations by using AI to uncover quantitative relationships between the complex 3D microstructure of materials and their resulting effective macroscopic properties that are critical but have been underrepresented in traditional models. By integrating AI into traditional computational methods, our framework improves the accuracy and robustness of predictive equations. This integration represents a critical advance in materials science, providing new insights that are essential for designing and optimizing next-generation materials with tailored effective macroscopic properties. These materials are in increasing demand due to their critical role in various innovative applications, from energy storage in lithium-ion batteries to filtration systems in biomedical devices.

\smallskip
\noindent
Furthermore, the rapid development of porous media technology underscores the need for advanced predictive models that can adapt to complex conditions. Recent advances in material science and engineering have expanded the range of porous media applications and require a more sophisticated understanding of their microstructural influences on macroscopic properties. This complexity requires a multidimensional approach that predicts material behavior under controlled conditions and explores potential responses within defined parameters. Our framework addresses this need by incorporating multi-faceted data analysis and using computational advances to bridge the gap between theoretical models and real-world applications. These advances are crucial as they enable the development of more efficient, durable, and economically viable materials that are better able to meet the challenges of modern technological applications and environmental conditions.

\smallskip

\noindent
The rest of the present study is structured to first revisit the methodology used to generate and analyze the original dataset of 3D microstructures. Subsequent sections will discuss the implementation of the hybrid AI framework, present the results obtained from this analytical approach, and compare these results with those of the initial study to highlight improvements and new findings. Finally, we will explore the implications of these results for future research and the development of novel material designs.

\section{Materials and methods}

\subsection{Data overview}

In this study, we use the comprehensive dataset of 90,000 3D microstructures previously generated and characterized in \cite{prifling2021}, representing a wide array of microstructures of porous media generated by means of  various spatial stochastic models. In particular, nine types of stochastic 3D models have been used in the original study, each of which has been deployed to generate 10,000 unique microstructures. This extensive dataset was designed to uniformly cover porosity values $\varepsilon$ between 0.3 and 0.95, while also spanning a wide range of other transport-relevant structural descriptors as well as macroscopic properties like effective diffusivity.

\begin{figure}[h!]
    \centering
    \includegraphics[width=0.7\linewidth]{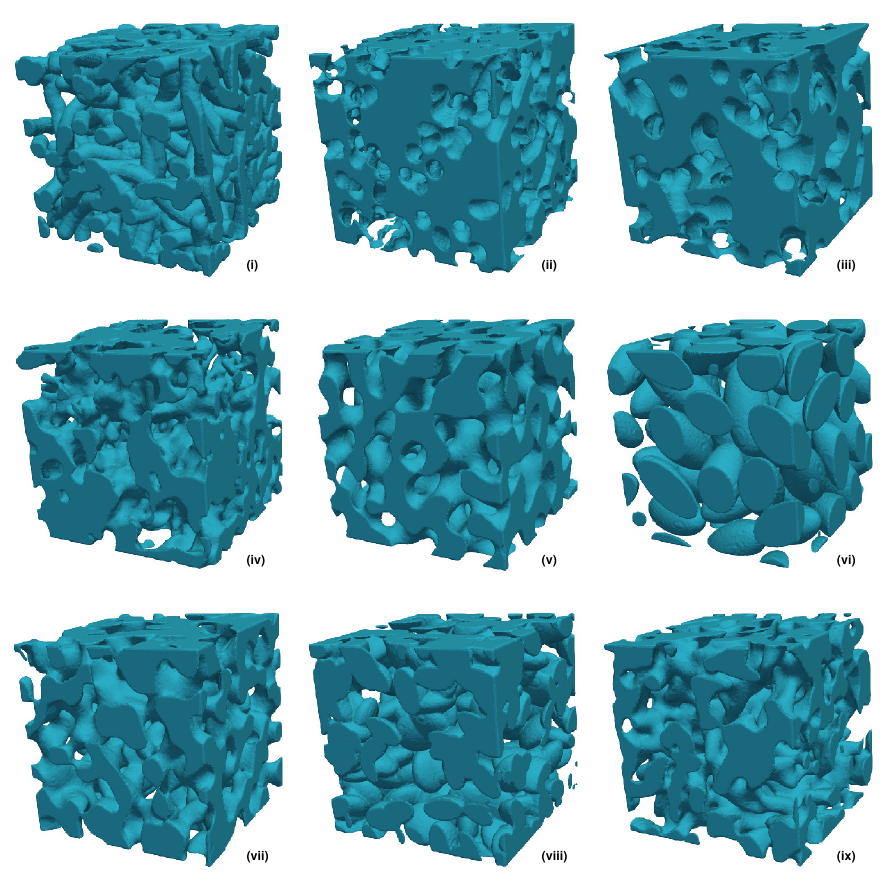}
    \caption{3D renderings of simulated structures, where transport takes place in the transparent pore space. The structures correspond to the nine different types of stochastic microstructure models, namely fiber system (I), channel system (II), spatial graph (III), level set of a Gaussian random field (IV), level set of a spinodal decomposition (V), as well as systems of hard ellipsoids (VI), smoothed hard ellipsoids, (VII), soft ellipsoids (VIII), and smoothed soft ellipsoids (IX). The figure is reproduced from \cite{prifling2021}.}
    \label{fig:structures}
\end{figure}

\smallskip
\noindent
It is important to highlight that the parameters of the stochastic 3D models have been varied systematically, such that the virtual 3D microstructures exhibit numerous different structural scenarios that roughly ensure uniformity of the values of effective diffusivity (called $M$-factor in the following). Note that the virtual 3D microstructures have been generated on a discrete cuboidal sampling window consisting of $192^3$ voxels with periodic boundary conditions, which standardizes the analysis and comparison of different stochastic 3D models. Moreover, the stochastic 3D models encompass a variety of structures, including random fiber systems \cite{townsend.2021}, random channel systems arising by considering the complement of fiber systems, spatial stochastic graphs \cite{gaiselmann.2014}, level sets of Gaussian random fields \cite{lantuejoul.2013, chiu.2013}, level sets of spinodal decompositions by numerically solving the Cahn-Hilliard equation \cite{miranville.2019}, and several types of random ellipsoid configurations including ellipsoids smoothed with a Gaussian filter \cite{roeding.2017, roding.2018, gonzalez.2008}, see Figure~\ref{fig:structures}.

\noindent
More precisely, we utilize the dataset described in \cite{prifling2021} consisting of several structural descriptors as well as two effective macroscopic properties (namely effective diffusivity and permeability), which have been computed for $90,000$ virtually generated porous microstructures. Specifically, the dataset is represented as follows:
\begin{equation}
    \mathcal{A} = \left(
        (\varepsilon_i,S_i, m(\tau_\mathrm{geo})_i,  \sigma(\tau_\mathrm{geo})_i, \beta_i, M_i)
    \right)_{i=1,\dots,n},
\end{equation}
where $n=90,000$ is the number of entries. The structural descriptors of the porous microstructures include porosity $\varepsilon$, specific surface area $S$, mean value $m(\tau_\mathrm{geo})$ and standard deviation $\sigma(\tau_\mathrm{geo})$ of geodesic tortuosity as well as constrictivity $\beta$ of the pore space. Further details regarding the formal definition of these microstructural descriptors and their estimation from voxelized 3D image data can be found in \cite{prifling2021}. Furthermore, the entries in the dataset $\mathcal{A}$ comprise an effective macroscopic property, namely, the $M$-factor, which is defined as the ratio of effective and intrinsic diffusivity. Various studies have shown, that the considered structural descriptors influence the $M$-factor and postulated various empirically derived equations \cite{prifling2021, gaiselmann.2014, prifling.2023soil}.

\smallskip
\noindent
The diversity of the dataset $\mathcal{A}$ facilitates a detailed exploration of the intrinsic relationships between microstructure characteristics and material properties, providing a robust foundation for the application of our hybrid AI framework in the current study. This framework, described in the subsequent section, aims to derive precise and interpretable equations that model the quantitative relationships between structural descriptors and effective macroscopic properties. Using this framework, we enhance our ability to predict material properties based on microstructural characteristics, significantly advancing predictive modeling capabilities in materials science.

\subsection{Hybrid AI framework implementation} \label{sec:Framework}

A custom-developed hybrid AI framework is employed to investigate the structure-property relationship, linking structural descriptors of 3D microstructures—such as porosity, specific surface area, and geodesic tortuosity—to their effective transport property. This framework integrates various computational and data processing modules including dimensional analysis (DA), symbolic regression (SR), transformation techniques (TTs), and deep learning networks. Collectively, these modules enhance the framework’s efficiency through systematic pre-processing, optimization, and execution phases, ultimately enabling the derivation of robust and interpretable analytical equations. 

\smallskip
\noindent
\textbf{Overview of the hybrid AI framework.} This section provides a general overview of our hybrid AI framework, which integrates multiple analytical modules to model structure-property relationships within varied datasets.\cite{HOSSEINHASHEMI2023} Initially, we outline the general workflow of the framework and provide a brief overview of all integrated modules and their role in improving our modeling capabilities. Following this overview, we delve specifically into one of the modules of our framework, the graph attention network (GAT), which was particularly instrumental in this case study. We detail the applications of GAT and the enhancements made to optimize its performance within our framework.

\smallskip
\noindent
\textbf{Data acquisition and processing.}
Within this framework, the workflow begins by entering the data into the DA module \cite{udrescu2020aifeynman}. The input data comprises structural descriptors of 3D microstructures, which include features such as porosity, specific surface area or mean geodesic tortuosity. These are crucial for modeling the physical characteristics of the structures. The output data, represented as the transport property, include properties such as diffusivity or permeability, depending on the specific transport phenomena being studied. These data are processed with their corresponding SI units to determine "dimensionless groups" that ensure the physical validity of the derived equations. These dimensionless groups are sets of related features --- where "features" refer to the microstructural descriptors --- that combine through specific mathematical operations to form dimensionless quantities, reflecting intrinsic properties of the system that are invariant across different scales. The original dataset is updated by incorporating these dimensionless groups into the dataset, which improves the ability of the dataset to represent physical relationships more comprehensively. Specifically, the update process involves adjusting the dataset by adding new columns for each discovered dimensionless group, allowing the framework to utilize these groups in subsequent modules to refine input data and improve the accuracy of the final derived equation. This updated data set then progresses through the framework, where further analyses are performed to deepen the understanding of structure-property relationships.

\smallskip
\noindent
\textbf{Symbolic regression (SR) and deep neural networks (DNN).}
After the DA module processes the data, the workflow transitions to the SR module, which includes several internal modules to develop precise mathematical equations describing input-output relationships. These internal modules use open-source software libraries such as PySR \cite{cranmer2023interpretable} and Gplearn \cite{stephens2016gplearn} alongside our in-house developed techniques. These techniques include the fragment selection technique (FST) \cite{HOSSEINHASHEMI2023,hosseinhashemi2024} and constant selection techniques (CST) \cite{HOSSEINHASHEMI2023,hosseinhashemi2024}, as well as the GAT for advanced data analysis.

\smallskip
\noindent
The SR module systematically works through the dataset to derive equations that effectively capture the relationship between inputs and output. If the initial results from the SR module do not meet our expectations, the workflow progresses to our in-house developed DNN \cite{HOSSEINHASHEMI2023} module. This DNN is specifically tailored to evaluate and refine physical equations provided by the user. It achieves this by predicting outputs and assessing their alignment with existing equations. The DNN performs this by generating new features from its predictions, which are then compared against the physical equations. If the comparison shows a close match, we make necessary adjustments to the equations, enhancing their accuracy and relevance. This iterative process involves breaking down complex equations, solving each component separately, and integrating them to build a comprehensive and accurate material model equation \cite{HOSSEINHASHEMI2023}. 

\smallskip
\noindent
The iterative process within the framework is characterized by a continuous refinement cycle. This cycle, powered by a feedback loop, rigorously evaluates the equations' effectiveness. If an equation does not meet our criteria for accuracy and utility, the data is reprocessed with newly updated features and parameters for another iteration. This cycle repeats until an equation that best fits both the empirical data and theoretical understanding is identified and chosen as the final output for the framework. It is important to note that the detailed descriptions of these modules and their specific configurations are thoroughly explained in \cite{HOSSEINHASHEMI2023, hosseinhashemi2024}, ensuring that the focus here remains on the general workflow and new components, such as the GAT, rather than repeating intricate technical details of the entire framework with the different previously developed modeling modules.

\begin{figure}[h] 
    \centering
   \includegraphics[width=0.99\linewidth]{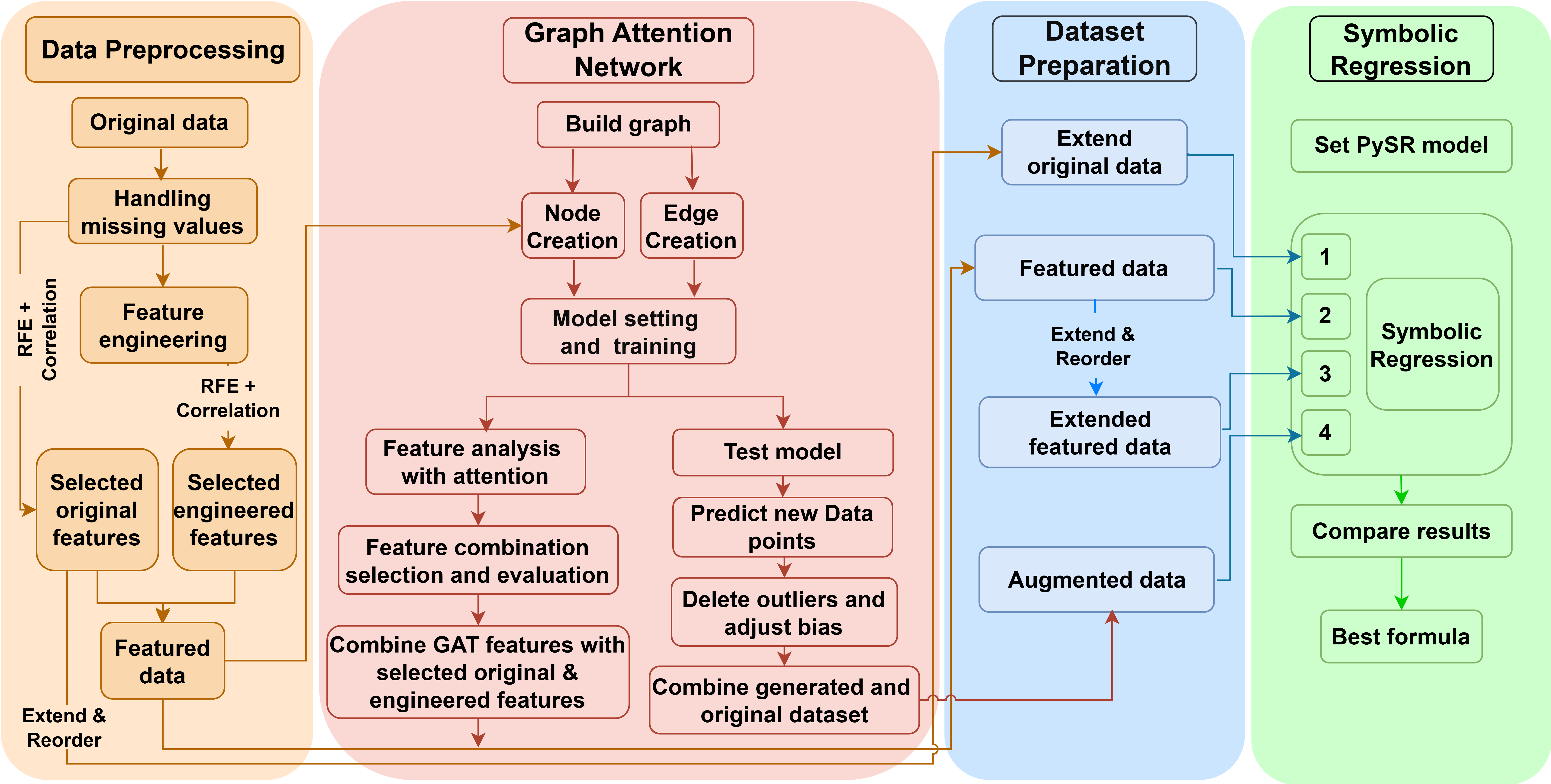}
    \caption{Graph attention network (GAT) Workflow in the hybrid AI framework, showing the four sub-modules: Data pre-processing, Graph attention network, Dataset preparation, and Symbolic regression, which collectively enhance data analysis and model predictions.}
    \label{fig:workflow}
\end{figure}

\smallskip
\noindent
\textbf{Integrating the graph attention network (GAT) to enhance our framework's capabilities.} 
\noindent
In our hybrid AI framework, various modules are evaluated to determine which one delivers the best equation for the structure-property relationships in specific case studies. We discovered that the GAT module was particularly effective for this dataset due to its capability to manage complex data relationships. Consequently, our efforts in this study focused not only on advancing the GAT module as a method for data augmentation—as previously detailed in \cite{hosseinhashemi2024}—but also on investigating the interactions between input features. Furthermore, we refined the GAT workflow by incorporating pre-processing methods such as “feature selection” with recursive feature elimination (RFE) \cite{barzani2024evaluating,jeon2020hybrid}, correlation analysis, and feature engineering \cite{khasne2025educational}. Given the extensive utilization and enhancement of the GAT within this study, we provide a detailed description here. Readers interested in exploring other modules utilized within our framework can refer to references \cite{HOSSEINHASHEMI2023, hosseinhashemi2024} for additional insights. As shown in Figure~\ref{fig:workflow}, the GAT module comprises four interconnected sub-modules. These include:
(i) Data pre-processing sub-module: Optimizes the initial data setup,
(ii) Graph attention network sub-module: Focuses on identifying and analyzing combinations of features and enhances data augmentation,
(iii) Dataset preparation sub-module: Handles the integration and organization of data, and
(iv) Symbolic regression sub-module: Responsible for generating the final mathematical equations.

\noindent
Each sub-module is designed to address specific regression challenges, and their collaborative operation ensures efficient and seamless data processing throughout the workflow. Data transitions between the sub-modules in a structured manner, allowing for systematic handling and transformation of information according to defined protocols. In the following sections, each sub-module will be discussed in detail to provide a comprehensive understanding of their functions and contributions to the overall process.

\begin{enumerate}
    \item [(i)] \textbf{Data pre-processing sub-module.} 
    \hangindent=2em
    The workflow of the GAT module begins with the "Data Pre-processing" sub-module, which initially addresses raw data by handling missing values. Here, we employ the mean padding approach, which replaces missing values (NaNs) with the mean of the available data for each feature. This approach is computationally efficient, helps prevent the introduction of extreme outliers, and preserves the overall statistical distribution of the dataset—critical factors for successful SR tasks \cite{alwateer2024missing}.
    
Following the initial data cleaning, this sub-module introduces a dual feature selection mechanism that combines recursive feature elimination (RFE) and correlation analysis to identify and retain the most representative features from the original dataset, termed “Selected original features.” The dual mechanism works by first using RFE, where a model, such as a RandomForestRegressor, iteratively removes the least important features based on their impact on model accuracy. Simultaneously, correlation analysis assesses the strength and direction of the linear and nonlinear relationships between each feature and the target variable. Throughout this process, each feature is assigned an importance score that quantifies its effectiveness in predicting the target variable. This scoring ensures that only those features that demonstrate significant relevance and strong correlation are retained. This dual selection mechanism aims to refine the feature set by systematically removing the least important features and effectively filtering the dataset to include only the most relevant features strongly linked to the target variable.

Following this selection, the workflow moves into the feature engineering method, where input features are enriched and refined. This method generates new features by combining existing ones through basic operations such as addition, subtraction, multiplication, and division. Additionally, more advanced transformations inspired by physical laws are applied. These transformations are methodically organized into four levels: basic binary operations (e.g., addition and multiplication); interactions among three features (e.g., combinations like multiplication or ratios of three features); power transformations (e.g., squares, cubes, and inverses); and polynomial feature generation using libraries like sklearn's PolynomialFeatures. These methods are strategically designed to capture both simple and complex relationships between features while ensuring the physical significance of the features is preserved. 

After feature engineering, the entire feature set — including both newly engineered and original features — undergoes the same rigorous dual selection mechanism introduced earlier. The features that successfully pass through this dual selection mechanism are termed "Selected engineered features," distinguished by their relevance and non-redundancy in representing the underlying data patterns. This mechanism ensures that only the most relevant features are retained, effectively balancing computational efficiency with data quality.

The "selected original features" and "selected engineered features" are then merged to form a comprehensive dataset, referred to as "Featured data." This Featured data is subsequently passed to the “Dataset preparation” and the “Graph attention network” Sub-module for further analysis, ensuring that the data is thoroughly prepared for the subsequent stages of the workflow.

\item [(ii)]\textbf{Graph attention network sub-module.} 
    \hangindent=2em
    The "Graph attention network" sub-module employs the GATv2 model \cite{brody2022how} for two specific and distinct purposes: advanced data augmentation and enhanced feature interaction analysis. The selection of GATv2 over the original GAT model \cite{velickovic2018graph} is based on its superior capability in handling dynamic attention, allowing for attention weights to adapt flexibly across different nodes. This characteristic is crucial for effectively capturing complex feature interactions, which is central to both data augmentation and in-depth feature analysis.  
    
\smallskip
\noindent
\textbf{First use: Advanced data augmentation.} Initially, the GATv2 model is utilized for data augmentation, as elaborated in our prior publication \cite{hosseinhashemi2024}. This involves constructing a dense graph where each data point is represented as an interconnected node. The model is trained on this graph, dividing the data into an 80\% training set and a 20\% testing set to enhance predictive accuracy and generalizability. An early stopping mechanism during training halts the process if model performance declines below a set threshold, with subsequent retraining efforts to restore performance levels. Additionally, we implement a filtering step based on mean absolute error to refine both the training and testing sets by removing highly erroneous predictions. Once training is complete, the model generates new data points by applying learned complex relationships, which are then merged with the original dataset to create "Augmented data." This enriched dataset enhances the subsequent symbolic regression analysis, improving both accuracy and generalizability.

\smallskip
\noindent
\textbf{Second use: Enhanced feature interaction analysis.} Expanding on the concepts detailed in \cite{hosseinhashemi2024}, we further utilize the GATv2 model to deepen our understanding of underlying data patterns and refine our data augmentation processes. The framework constructs a sophisticated graph structure, representing each data sample as a node. Relationships between nodes are defined using cosine similarity, which effectively captures directional similarities between feature vectors—preferred over euclidean distance to better handle the complexities of high-dimensional spaces \cite{Goodfellow-et-al-2016}. In this graph, nodes represent the values of all selected features, normalized to ensure numerical stability and expedite model convergence. For the data augmentation method, nodes encompass the entire feature set, and edges are fully connected to maximize information flow. In contrast, for enhanced feature interaction analysis, the graph construction is more nuanced. Nodes are standardized by feature values, and edges are selectively established based on cosine similarity. Edge attributes, derived from similarity scores, provide a detailed structure for analyzing feature interactions.

The GATv2 model employs a multi-head attention mechanism, where each layer dynamically adjusts the number of heads to explore feature interactions at various levels. This mechanism enables nodes to focus on neighboring nodes with calculated attention scores, enhanced by a shared linear transformation, LeakyReLU activation, and softmax normalization \cite{vaswani2017attention}. Attention weights are dynamically updated during the message-passing phase, improving the model’s ability to analyze feature relationships from multiple perspectives and perform regression tasks effectively. The results from this feature interaction analysis, including important binary and ternary feature combinations, are referred to as 'GAT features' within the framework. These “GAT features” are then integrated with “selected original” and “engineered features”, forming enriched datasets for advanced analytical techniques. This robust integration sets a solid foundation for the dataset preparation module, enhancing the accuracy and generalizability of the symbolic regression results.

\item [(iii)]\textbf{Dataset preparation sub-module.}
    \hangindent=2em
    The “Dataset preparation” sub-module combines data from multiple sources: “Selected original features”, “Featured data”, and “Augmented data”. This integration prepares four types of datasets:

    \begin{itemize}
    \item \textbf{Extended original dataset:} Created by duplicating and randomly rearranging the "Selected original features." This process is initiated by setting the “extend\_time” parameter, ensuring that the statistical properties of the data are maintained to enhance the dataset's representativeness and improve the model’s ability to generalize across different scenarios.
    
    \item \textbf{Featured dataset:} Comprising both original, engineered, and GAT features, this dataset is refined through methods like feature engineering and the GAT module. It also undergoes a filtering mechanism using RFE and correlation analysis to ensure only the most impactful features are included. The dataset is further assessed to maintain mathematical correctness, preventing computational errors such as division by zero or incorrect power calculations.

    \item \textbf{Extended featured dataset:} This dataset includes more extensive versions of the “Featured dataset”. It is produced by extending the dataset when the “extend\_time” parameter is set above a threshold, which serves to increase the sample size and enhance the robustness and stability of the dataset for a more reliable modeling outcome.

    \item \textbf{Augmented dataset:} This dataset incorporates new data points generated by the GATv2 model, enriching the dataset with augmented samples. This integration not only adds diversity to the data but also enhances the dataset’s complexity, allowing for more nuanced analysis and modeling in the symbolic regression phase.

     \end{itemize}

Mechanisms are implemented to handle potential computational issues, such as verifying divisors and checking power operation ranges to maintain numerical stability during feature combination evaluation. This ensures that the feature combinations remain valid and usable for downstream tasks. The expanded datasets are formatted and saved as CSV files for traceability and smooth integration into subsequent SR modules. By employing dataset extension strategies and systematically combining data sources, the module ensures the availability of high-quality training datasets. This process effectively addresses challenges such as insufficient data and data sparsity while preserving the integrity and diversity of the original distribution.

\item [(iv)]\textbf{Symbolic regression sub-module.}
    \hangindent=2em
    The “Symbolic regression” sub-module is the concluding step of our framework. SR a machine learning method, attempts to find mathematical equations that effectively capture the relationships between input features and output targets. In this study, we incorporate the open-source PySR library \cite{cranmer2023interpretable} as our tool for SR. It employs genetic programming principles to evolve potential mathematical expressions through a structured, iterative process. This implementation begins by generating a diverse initial population of expressions and managing their complexity to foster a balanced search space. The library evaluates each expression's performance based on prediction accuracy and structural simplicity, employing metrics like mean squared error (MSE) and the Bayesian information criterion (BIC). The latter helps in selecting expressions that are both precise and predictive, avoiding overfitting. Key operations within this process include crossover and mutation—techniques that respectively recombine and alter expressions to explore new solutions. The output of the SR module is the optimal mathematical expression that best represents the relationships in the data. 

\end{enumerate}
\noindent
Comprehensive validation mechanisms are applied to ensure data consistency throughout the workflow. For example, feature-engineered datasets are validated before graph construction, and outputs from the GATv2 module are checked for completeness and accuracy before entering SR. Additionally, a logging system tracks data flow, intermediate results, and key parameters at every stage, facilitating traceability, debugging, and optimization. These mechanisms ensure the framework's reliability, efficiency, and seamless integration across all modules, culminating in high-quality SR results. The performance of the framework and the fit of the predicted models to the original data are evaluated using the mean absolute error (MAE), mean squared error (MSE), and the coefficient of determination ($R^2$), shown by the equations below:

\begin{equation}
    \text{MAE} = \frac{1}{n} \sum_{i=1}^n \left| y_{\text{pred}, i} - y_{\text{MPM}, i} \right|
\end{equation}

\begin{equation}
    \text{MSE} = \frac{1}{n} \sum_{i=1}^n \left( y_{\text{pred}, i} - y_{\text{MPM}, i} \right)^2
\end{equation}

\begin{equation}
    R^2 = 1 - \frac{\sum_{i=1}^n (y_{\text{pred}, i} - y_{\text{MPM}, i})^2}{\sum_{i=1}^n \left( y_{\text{pred}, i} - \frac{1}{n} \sum_{j=1}^n y_{\text{MPM}, j} \right)^2}
    \label{eq:r2}
\end{equation}
\noindent
Here, \(y_{\text{pred},i}\), \(y_{\text{MPM},i}\) denote the predicted and MPM (Model Predictive Maintenance) values of the output, respectively.  

\subsection{Construction of synthetic datasets for framework validation}\label{sec:statisticalAnalysis}

Building upon the methodologies described in Section~\ref{sec:Framework}, this section outlines how we validate the robustness of our hybrid AI framework. Specifically, we demonstrate the capabilities of the framework through its derived equation, which links the structural descriptors of dataset $\mathcal{A}$ to the $M$-factor. This validation is crucial for ensuring the practical applicability of the predictive equations, especially under real-world conditions where data may not be ideal. Through this approach, we not only deepen our understanding of the microstructure-property relationships but also enhance the robustness of our predictive equations. The core of our validation process involves Eq.~(10) of \cite{prifling2021}, which is as follows:
\begin{equation}\label{eq:M_assumption}
    M(\varepsilon,S, m(\tau_\mathrm{geo}),  \sigma(\tau_\mathrm{geo}), \beta ) = \frac{\varepsilon}{m(\tau_\mathrm{geo})^\alpha},
\end{equation}
with $\alpha = 8.483$. Note that this type of equation has been originally considered in \cite{stenzel.2016} and later reconsidered in \cite{neumann.2020}. Although the derivation and implications of this equation are discussed in detail in Section~\ref{sec:statistical:analysis}, it is integral to the methodology described here as it forms the basis for our simulation studies. To quantitatively assess the reliability of the framework’s result described in Section ~\ref{sec:statistical:analysis}, we utilize the structural descriptors within dataset $\mathcal{A}$ together with the formula given in Eq.~\eqref{eq:M_assumption}, to generate new datasets with varying noise levels for the output variable $M$-factor. More precisely, we perform a simulation study for which we assume that Eq.~\eqref{eq:M_assumption} relates structural descriptors $(\varepsilon_i,S_i, m(\tau_\mathrm{geo})_i, \sigma(\tau_\mathrm{geo})_i, \beta_i)$ with the $M$-factor. Therefore, we want to assign to each descriptor vector $x_i= (\varepsilon_i,S_i, m(\tau_\mathrm{geo})_i, \sigma(\tau_\mathrm{geo})_i, \beta_i)$ a randomly generated value for the $M$-factor, where the generation upholds the following conditions:
\begin{itemize}
    \item[(i)] The sampled $M$-factor values are on average $M(x_i)$, i.e., conditioned on $x_i$ the mean value of the $M$-factor distribution follows Eq.~\eqref{eq:M_assumption}.
    \item[(ii)] The sampled $M$-factor values have a standard deviation of $\noise$.
    \item[(iii)] The sampled $M$-factor values lie within the interval $[0,\varepsilon_i]$, as the $M$-factor is non-negative and bounded by the porosity \cite{Torquato.2002}.
\end{itemize}
Therefore, to uphold (iii), we utilize a truncated normal distribution, the probability density of which is given by
\begin{equation}
    f_{\mu_\mathrm{G},\sigma_\mathrm{G},\varepsilon_i}(z) = 
    \left\{
    \begin{array}{ll}
       \frac{1}{
        \Phi((\varepsilon_i-\mu_\mathrm{G})/ \sigma_\mathrm{G})  - \Phi(-\mu_\mathrm{G}/ \sigma_\mathrm{G}) 
        }
        \frac{1}{\sqrt{2 \pi \sigma_\mathrm{G}^2}}
        \exp\!\left( - \frac{(z-\mu_\mathrm{G})^2}{2 \sigma_\mathrm{G}^2} \right), 
          &  \text{if } 0\leq z \leq \varepsilon_i, \\
        0, & \text{else,}
    \end{array}
    \right.
\end{equation}
where $\Phi:\R\to[0,1]$ denotes the cumulative distribution function of the standard normal distribution and $\mu_\mathrm{G}\in\R$ and $\sigma_\mathrm{G}>0$ are the mean value and standard deviation of the underlying untruncated normal distribution, which do not necessarily coincide with the actual mean value and standard deviation of the truncated distribution. Thus,  $\mu_\mathrm{G}$ and $\sigma_\mathrm{G}$  have to be calibrated such that the conditions (i) and (ii) are satisfied, i.e., such that the mean value and standard deviation of $f_{\mu_\mathrm{G},\sigma_\mathrm{G},\varepsilon_i}$ are given by $M(x_i)$ and $\sigma_\mathrm{noise}$ for a given structural descriptor $x_i$. Therefore, we solve the optimization problem
\begin{equation}
\begin{aligned}
(\mu_{\mathrm{G},i},\sigma_{\mathrm{G},i}) = &\argmin_{(\mu_\mathrm{G},\sigma_\mathrm{G}) \in (0,\infty)^{2}}  \Biggl( \biggl| M_i - \int_{0}^{\varepsilon_i} z f_{\mu_\mathrm{G},\sigma_\mathrm{G},\varepsilon_i}(z) \, \mathrm{d}z \biggr| \\ 
&+ 
   \biggl| \sigma_\mathrm{noise} - \sqrt{
    \int_0^{\varepsilon_i}
    \Bigl(z - \int_0^\varepsilon z^\prime f_{\mu_\mathrm{G},\sigma_\mathrm{G},\varepsilon_i}(z^\prime)\,\mathrm{d}z^\prime \Bigr)^2 f_{\mu_\mathrm{G},\sigma_\mathrm{G},\varepsilon_i}(z)\, \mathrm{d}z
   } \ \biggr| \ \Biggr),
\end{aligned}
\end{equation}
where $\mu_{\mathrm{G},i},\sigma_{\mathrm{G},i}$ are the optimal choices for the parameters of $f_{\mu_\mathrm{G},\sigma_\mathrm{G},\varepsilon_i}$ for a given structural descriptor $x_i$.
Finally, for each structural descriptor $x_i$, we simulate a random number $M_{\mathrm{sim},i}$ according to its corresponding probability density $f_{\mu_{\mathrm{G},i},\sigma_{\mathrm{G},i},\varepsilon_i}$. Then, we construct a dataset $\mathcal{A}_{\sigma_{\mathrm{noise}}}$ of structural descriptors with known noise level of $\sigma_\mathrm{noise}$ by
\begin{equation}
    \mathcal{A}_{\sigma_\mathrm{noise}}=\bigl( (x_i, M_{\mathrm{sim},i})\bigr)_{i=1,\ldots,n}.
\end{equation}
To investigate the influence of increasing noise on interpretable equations determined by our AI hybrid framework, we repeated the construction of the generated datasets $\mathcal{A}_{\sigma_\mathrm{noise}}$ for noise levels $\sigma_\mathrm{noise} \in \{0.01, 0.02, \ldots, 0.16\}.$
To emphasize the key distinction between the noisy dataset $\mathcal{A}_{\sigma_\mathrm{noise}}$ and the original dataset $\mathcal{A}$, we highlight that in the case of $\mathcal{A}_{\sigma_\mathrm{noise}}$, the underlying functional relationship between the structural descriptors and the $M$-factor is explicitly known through Eq.~\eqref{eq:M_assumption}. In contrast, for the dataset $\mathcal{A}$ the exact form of the relationship is inferred from experimental or simulated data. Therefore, the controlled construction of $\mathcal{A}_{\sigma_\mathrm{noise}}$ allows us to quantitatively analyze how increasing noise levels impact the ability of the hybrid AI framework to recover interpretable equations and validate its robustness in extracting meaningful structure-property relationships.

\section{Results and discussion}

\subsection{Data pre-processing and feature importance analysis}
In the initial phase of our study, a comprehensive data pre-processing and feature importance analysis was conducted to determine the most influential factors affecting the transport property, $M$-factor. This step is critical as it sets the groundwork for applying our hybrid AI framework, ensuring that only the most relevant descriptors are carried forward to model development. We began by computing Pearson and Spearman correlation coefficients to analyze and identify linear and non-linear relationships between structural descriptors and the $M$-factor. Furthermore, during our analysis, additional engineered features such as $\varepsilon-m(\tau_\mathrm{geo})$ and $\varepsilon/m(\tau_\mathrm{geo})$ were found and emerged into the data through our feature engineering method. They were derived from the existing data to better capture the complex interactions and dependencies between descriptors that directly influence the value of $M$.

\smallskip
\noindent
The insights gained from these analyses were crucial in directing the next phase of our study—developing a predictive equation for $M$-factor. By reducing the dataset to the most influential features (descriptors) through feature engineering and selection processes, we enhanced both the robustness and interpretability of our equation. Understanding these relationships enabled us to apply our hybrid AI framework effectively, leading to the derivation of a precise and interpretable equation for $M$-factor, which we introduce in the following section. This work ensured that the transition from raw data analysis to refined predictive modeling was seamless and justified, providing a solid foundation for the high accuracy and robustness of the resulting equation.

\subsection{Development and validation of predictive equations}
Building on the insights from the pre-processing analysis, the predictive equation for the $M$-factor identified by our hybrid AI framework is presented in Table~\ref{tab:metrics}, alongside its corresponding performance metrics, including mean absolute error (MAE), mean squared error (MSE) and $R^2$.

\begin{table}[h!]
\centering
\caption{Identified equation for M-factor and the corresponding MAE, MSE, and $R^2$ metrics.}
\begin{tabular}{lcccc}
\hline
Identified equation & MAE & MSE & $R^2$ & Dataset \\ \hline
\multirow{2}{*}{$M = \dfrac{\varepsilon}{m(\tau_{\text{geo}})^{8.483}}$} & 0.0154 & $3.7 \cdot 10^{-4}$ & 0.992 & Train dataset \\
& 0.0167 & $3.8 \cdot 10^{-4}$ & 0.993 & Test dataset \\ \hline
\end{tabular}
\label{tab:metrics}
\end{table}

\begin{figure}[h!]
    \centering
    \begin{minipage}[b]{0.497\linewidth}
\includegraphics[width=\linewidth]{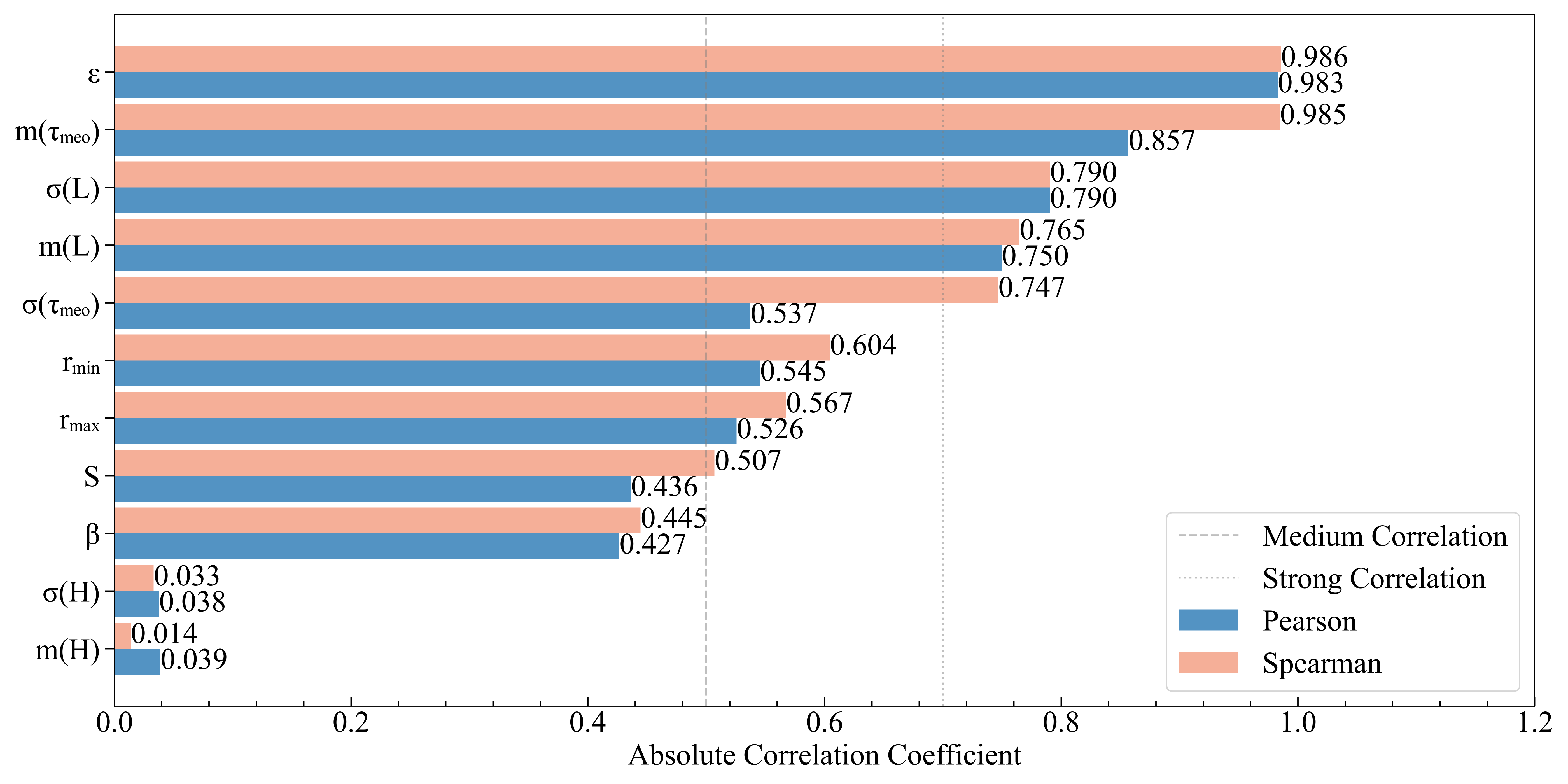}
        \label{fig:first-figure}
    \end{minipage}
    \hfill 
    \begin{minipage}[b]{0.497\linewidth}
\includegraphics[width=\linewidth]{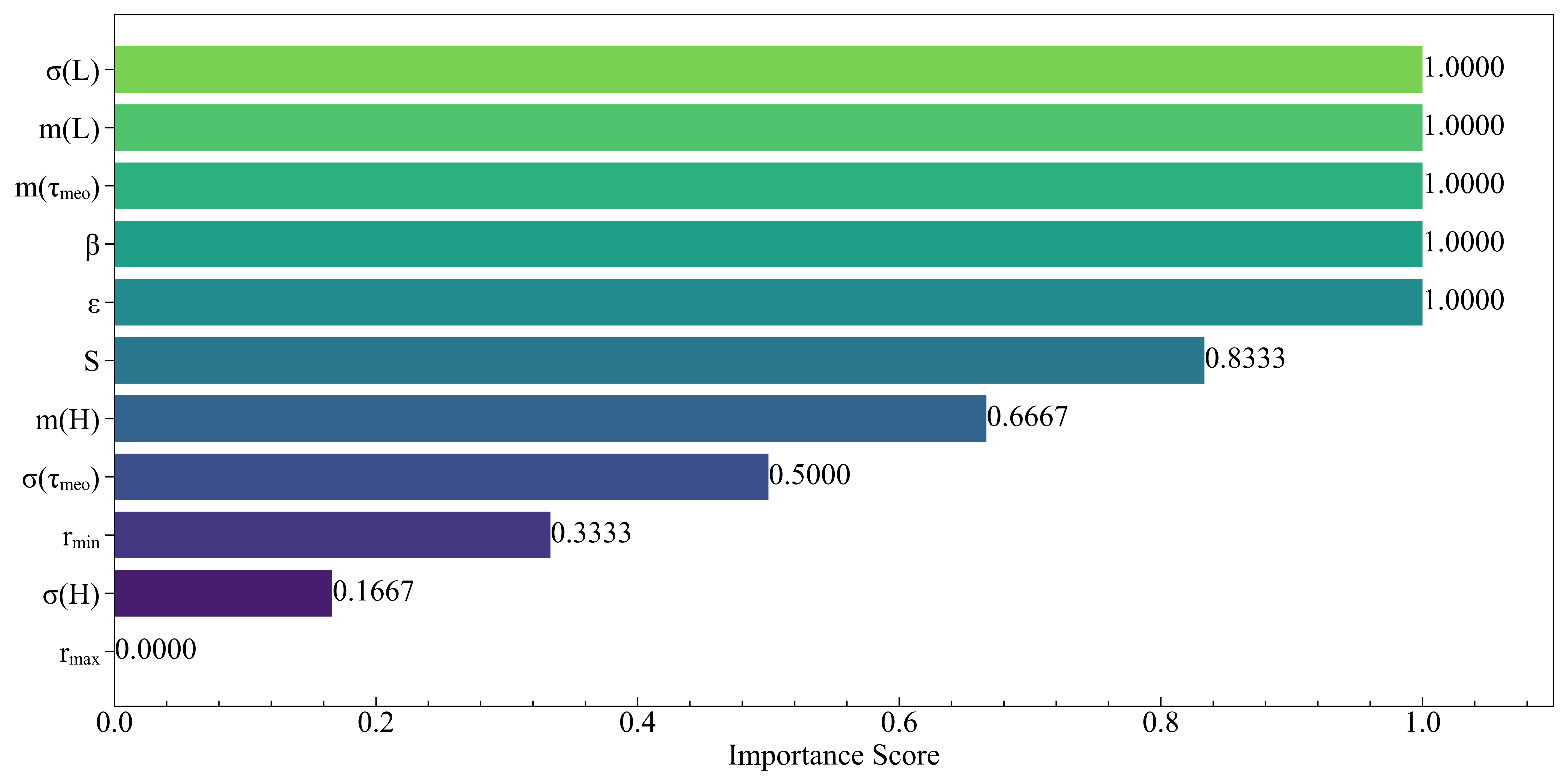}
        \label{fig:second-figure}
    \end{minipage}
    \caption{Correlation coefficients (left) and importance scores (right) of the features affecting $M$-factor, emphasizing the significant roles of porosity ($\varepsilon$) and geodesic tortuosity $(m(\tau_\mathrm{geo}))$.}
    \label{fig:correlationAndImportance}
\end{figure}

\noindent
This equation reaffirms the critical roles of porosity $\varepsilon$ and mean geodesic tortuosity $m(\tau_{\text{geo}})$ in determining the $M$-factor. By incorporating these key variables, the equation provides a comprehensive understanding of the descriptors influencing mass transport property in porous materials. The exponent $8.483$, fine-tuned through SR modules, reflects the nonlinear and nuanced impact of $m(\tau_{\text{geo}})$. The identified equation not only captures the most influential descriptors detected during our data pre-processing analysis (Figure \ref{fig:correlationAndImportance}) but also delivers an excellent fit to the data.

\begin{figure}[h]
    \centering
   \includegraphics[width=0.99\linewidth]{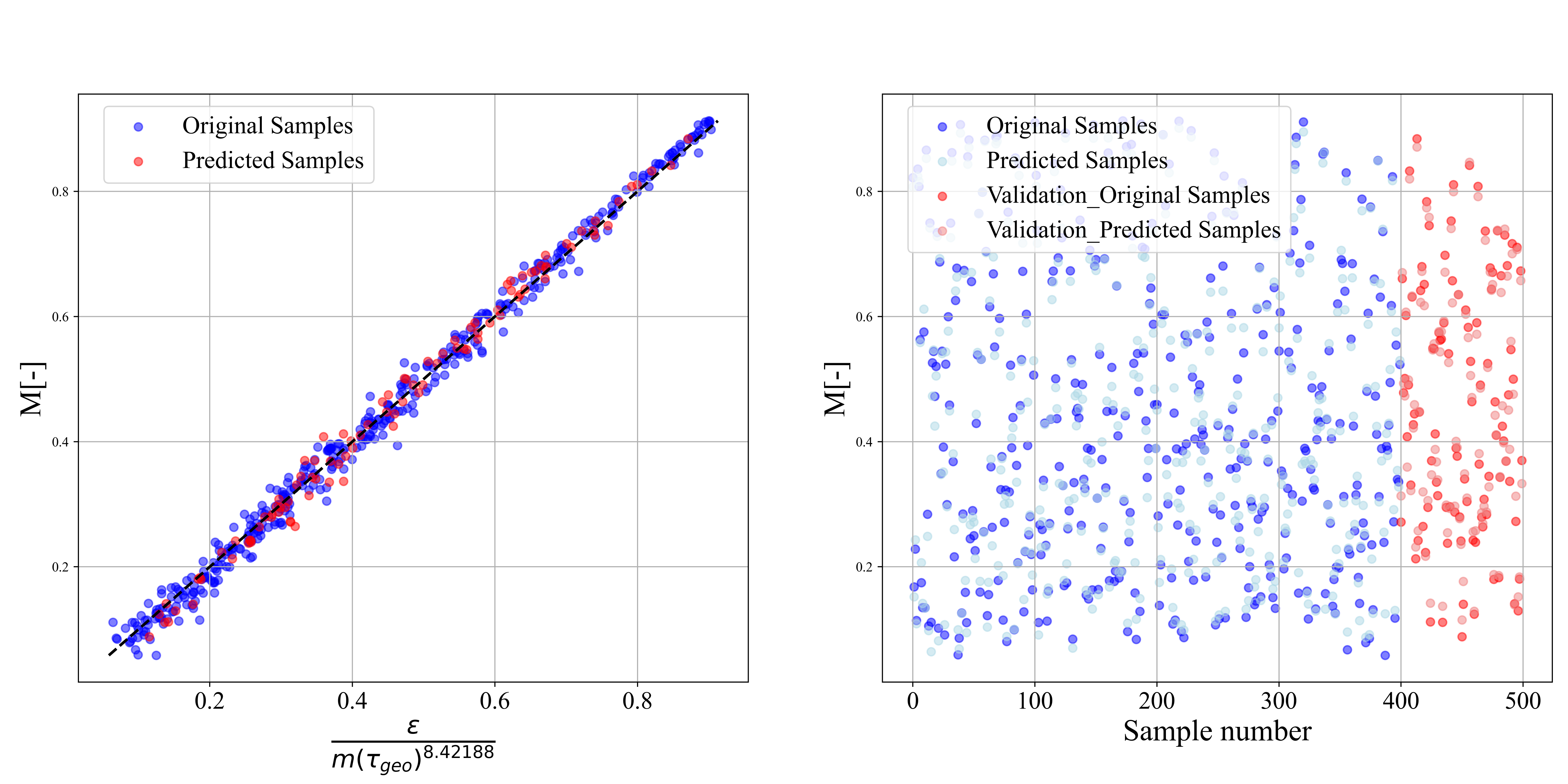}
    \caption{Actual $M$-factor vs. predicted $M$-factor (left) and a more detailed visualization (right) including the main samples (blue) and (right) validation samples (red). The corresponding performance metrics are given in Table~\ref{tab:metrics}.}
    \label{fig:ValidationMfactor}
\end{figure}

\noindent
To ensure the robustness and generalizability of the model, its performance was validated using both train and validation samples. Figure~\ref{fig:ValidationMfactor} illustrates the accuracy and reliability of the model across these datasets. In the left plot, the predicted values align closely with the actual values for $M$-factor, with data points from both training (80\%, in blue) and validation (20\%, in red) sets clustering along the diagonal line. This alignment, coupled with minimal residual scatter, demonstrates the model’s high accuracy and generalization capability. The evaluation metrics further underscores the equation's strong performance  listed in Table~\ref{tab:metrics}.

\smallskip
\noindent
The right plot further confirms the equation's consistency by visualizing predicted and actual values across all data points, separated into train and validation samples. The clustering of predicted values (training in dark blue and validation in dark red) around their respective actual values (lighter shades) highlights the equation’s reliability across different datasets. The absence of systematic deviation across these points reaffirms the robustness of the equation and its ability to generalize effectively. These results validate the success of our structured methodology, which began with a comprehensive exploratory analysis, followed by feature selection method like RFE and scatter plot evaluations. This process identified $\varepsilon$ and $m(\tau_\text{geo})$ as the most critical predictors. By leveraging our AI hybrid framework, we captured the complex nonlinear relationship between these descriptors and $M$-factor, ultimately yielding a simple yet powerful equation. The high $R^2$ scores and low error metrics across both main and validation samples confirm the equation’s accuracy and interpretability.

\subsection{Statistical analysis}\label{sec:statistical:analysis}

\noindent
In this section the performance of the hybrid AI framework is analyzed through a simulation study on the datasets introduced in Section~\ref{sec:statisticalAnalysis}. For this purpose, the $M$-factor of selected microstructures is determined using the equation: $M=\varepsilon/m(\tau_\text{geo})^{\alpha}$, where $\varepsilon>0$ denotes the volume fraction, $m(\tau_\text{geo})$ represents the mean geodesic tortuosity, and $\alpha=8.483$ is a constant. The calculated $M$-factor is then perturbed with truncated Gaussian noise characterized by a standard deviation $\noise$. The hybrid AI framework is applied $n$ times to the perturbed data, with increasing noise levels $\noise\in\{0.01,0.02,\ldots,0.16\}$. For each run $i=1,\ldots,n_{\text{run}}$ the hybrid AI framework identifies the three best equations that yield the highest regression performance based on the $\Rtwo$-score, as defined in Eq. \eqref{eq:r2}. Here, we have chosen $n_{\text{run}}=60$ to balance the number of data points with computational effort. The top three equations from each run $i \in \{1,\ldots,n_{\text{run}}\}$ of the hybrid AI framework are denoted as $M_{i,j},\ j=1,2,3$. The framework prepares four datasets from the original dataset to enhance our model's analysis capabilities, as detailed in the data pre-processing sub-module described in Section~\ref{sec:Framework}. However, since this study had sufficient data (90,000 data points), no dataset extension methods were applied, and the datasets were not expanded. Instead, the analysis was conducted on two dataset types: the "Augmented data" and the "Featured data", explained in section ~\ref{sec:Framework} (i) and (ii). Consequently, for each run, the hybrid AI framework generated the three most effective equations for these two dataset types.

\begin{figure}[htbp!]
    \centering
    \begin{minipage}[b]{0.39\linewidth}
        \includegraphics[width=\linewidth]{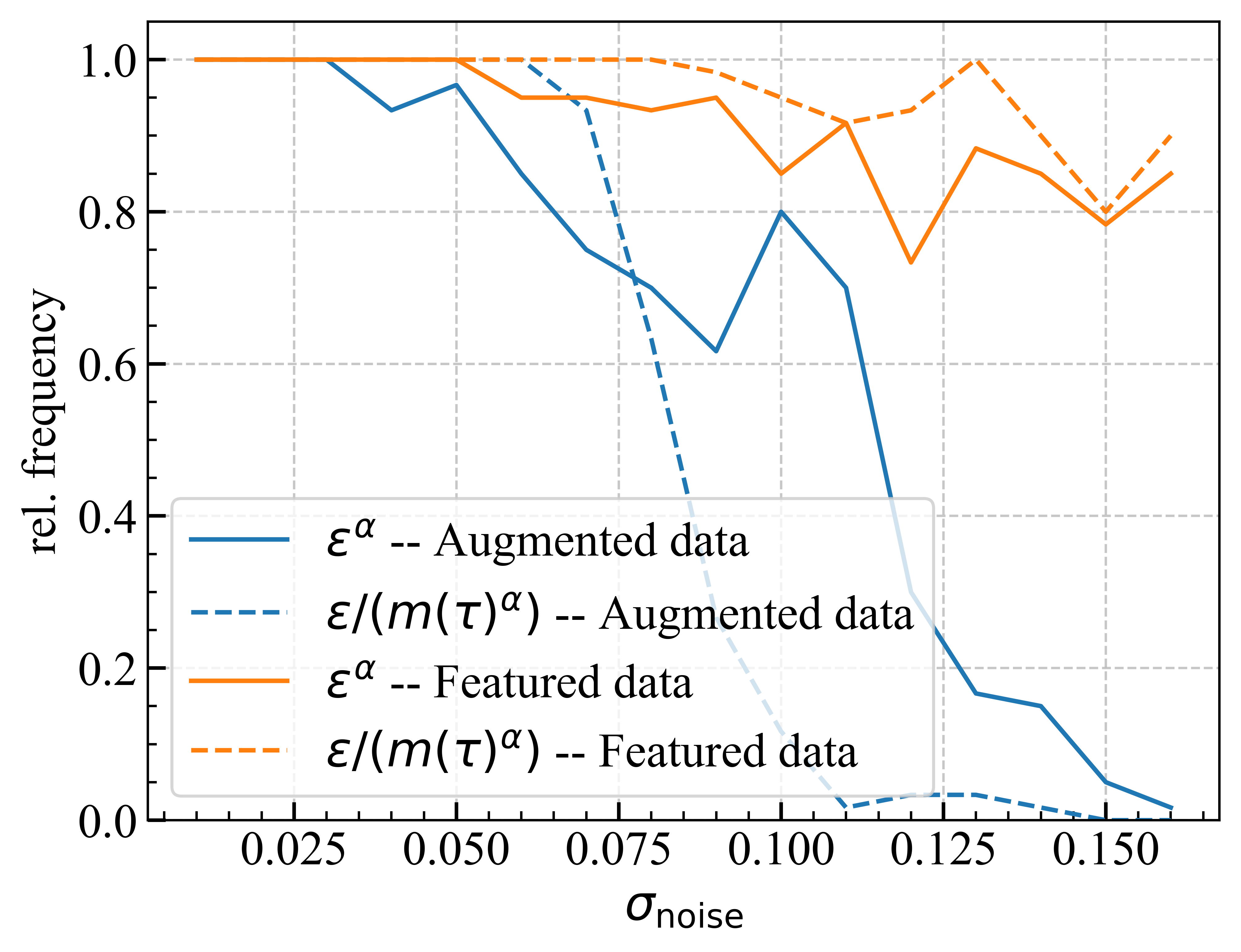}
    \end{minipage}
    \hfill 
    \begin{minipage}[b]{0.56\linewidth}
        \includegraphics[width=\linewidth]{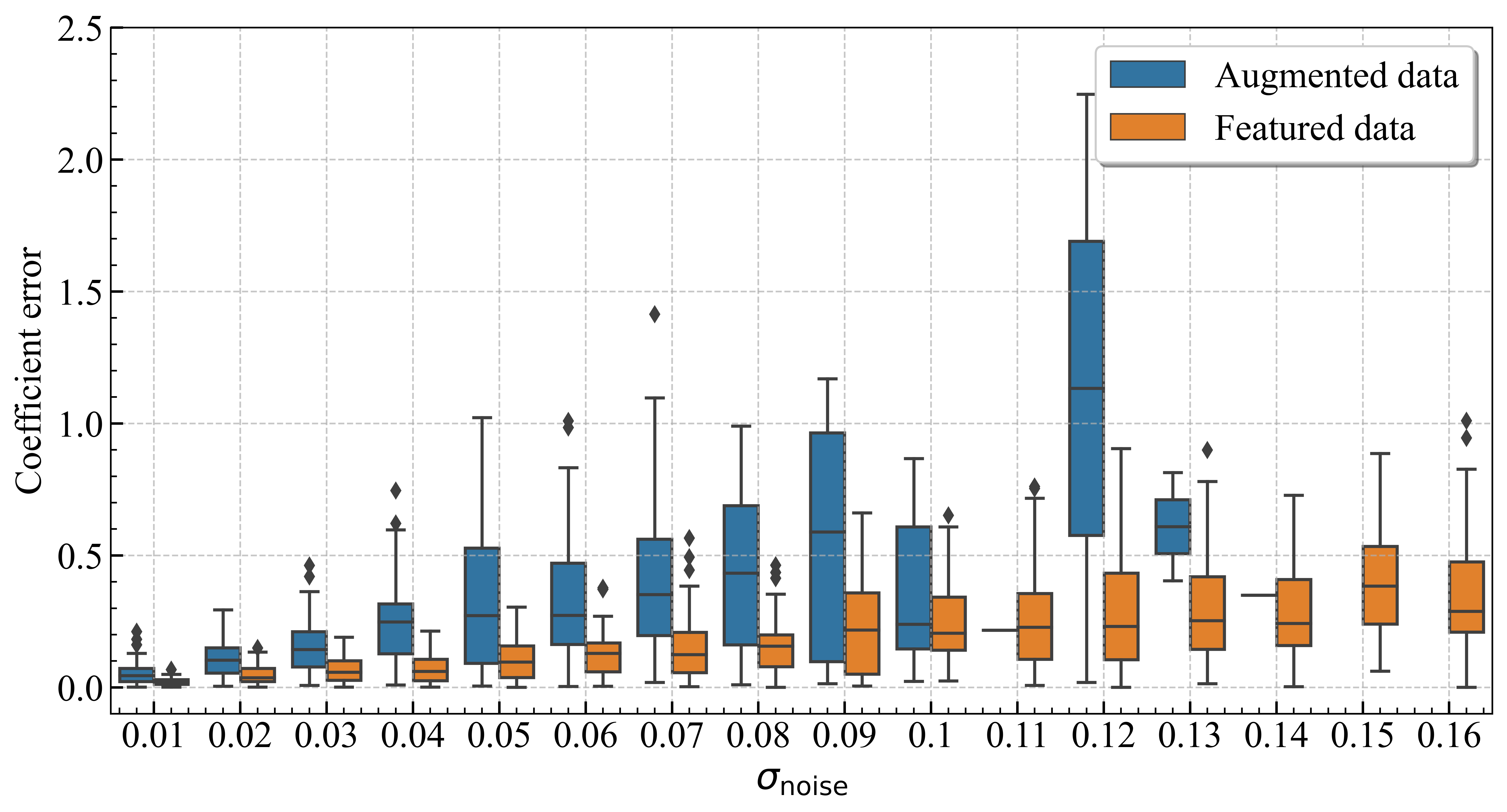}
    \end{minipage}
    \caption{Relative frequency (left) and coefficient error (right) of the true and alternative equations across noise levels for "GAT data" (blue) and "Featured data" (orange).}
    \label{fig:FrequencyAndCoeffError}
\end{figure}

\begin{figure}[htbp!]
    \centering   
    \begin{minipage}[b]{0.39\linewidth}
        \includegraphics[width=\linewidth]{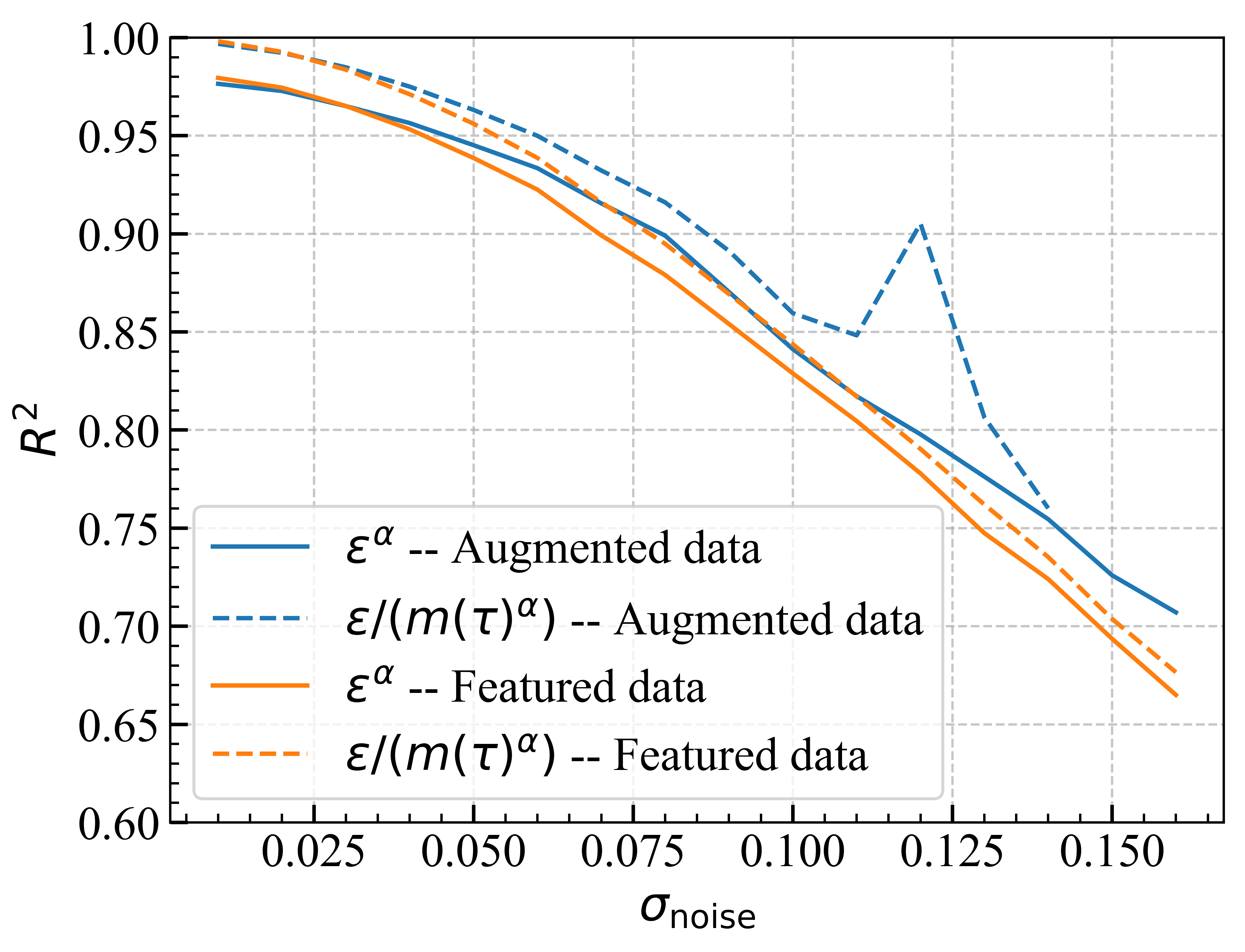}
    \end{minipage}
    \hfill  
    \begin{minipage}[b]{0.56\linewidth}
        \includegraphics[width=\linewidth]{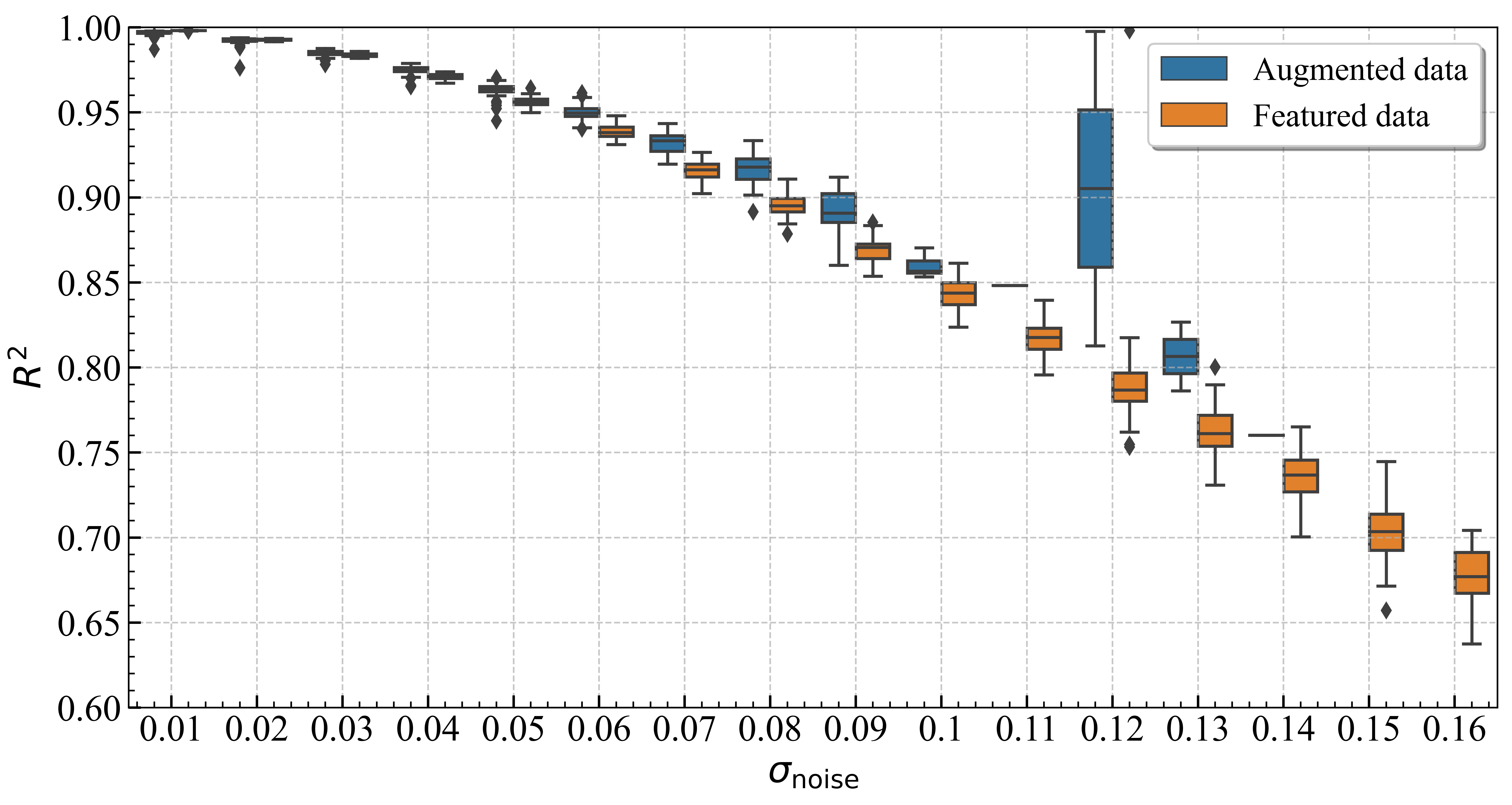}
    \end{minipage}
    \caption{
    (left) Mean $\Rtwo$-score and (right) $\Rtwo$-score distributions for the true ($\mPred=\varepsilon/m(\tau_\text{geo})^{\aPred}$, dashed lines/boxplots) and alternative ($\mPredPrime=\varepsilon^{\aPredPrime}$, solid lines) equations across noise levels, comparing "Augmented data" (blue) and "Featured data" (orange).}
    \label{fig:Rsquared}
\end{figure}

\smallskip
\noindent
Several performance metrics are determined to quantify the hybrid AI framework's performance. These metrics are calculated separately for each data type and each noise level $\noise$. The first metric, the relative frequency $f_M$, represents the likelihood that the correct equation type $\mPred=\varepsilon/m(\tau_\text{geo})^{\aPred}$ is identified among the top three regression equations. It is important to note that the hybrid AI framework also estimates $\aPred\in[0,\infty)$, which does not influence whether the equation $M_{i,j}$ determined by the hybrid AI framework is categorized as being from the correct or incorrect equation type. Mathematically, the relative frequency $f_M$ is expressed as:

\begin{align*}
    f_\mPred=\frac{1}{n}
    \bigl|\{i=1,\ldots,n \colon M_{i,j} \text{ is the correct equation type , for some } j \in \{1,2,3 \}\}\bigr| \in [0,1] \, ,
\end{align*}
where $|\cdot|$ denotes  cardinality. Alongside the correct equation type  $\mPred$, only the alternative equation $\mPredPrime=\varepsilon^{\aPredPrime}$ is predicted in a significant frequency. All other equation types are not considered in this analysis. The dependence of these relative frequencies $f_{\mPredPrime}$ of $f_{\mPred}$ on the noise level $\noise$ is illustrated in Figure~\ref{fig:FrequencyAndCoeffError} (left). Since the true equation for $M$ and the correct predicted equation type for $M_\text{pred}$ differ only by $\alpha$ and $\aPred$, boxplots indicating the distributions of $|\alpha-\aPred|$ are presented in Figure~\ref{fig:FrequencyAndCoeffError} (right). In Figure~\ref{fig:Rsquared} (left), the mean $\Rtwo$-scores, as defined in Eq.~\eqref{eq:r2}, of all predicted equation conditioned on $M_{i,j} = \mPred$ and $M_{i,j} =\mPredPrime$ are presented. Additionally, Figure~\ref{fig:Rsquared} (right) shows boxplots indicating the distribution of $\Rtwo$-score conditioned on the true equation $M=\varepsilon/m(\tau_\text{geo})^{8.483}$. Moreover, the (weighted) $\Lone$-error is defined as 
\begin{align*}
    \text{L}_1(M,\mPred)=\int_{\R^2_+}\bigl|M(x)-\mPred(x)\bigr|\ p(x)\ \text{d}x, 
\end{align*}
where $p(\cdot)$ denotes the probability density corresponding to the distribution of $x=(\varepsilon,m(\tau_\text{geo}))$.
Note, that although $M(\cdot)$ and $\mPred(\cdot)$ symbolize both the correct equation, they generally differ in their constants $\alpha$ and  $\aPred$, respectively. The $\Lone$-error is approximated by for the 90,000 virtually generated porous microstructures, introduced in \cite{prifling2021} by:

\begin{align*}
    \widehat{\text{L}}_1(M,\mPred)=\frac{1}{k}\sum_{i=1}^k|M(x_i)-\mPred(x_i)| \quad \text{ with } x_i=(\varepsilon_i,m(\tau_\text{geo})_i).
\end{align*}

\begin{figure}[h!]
    \centering
    \begin{minipage}[b]{0.39\linewidth}
        \includegraphics[width=\linewidth]{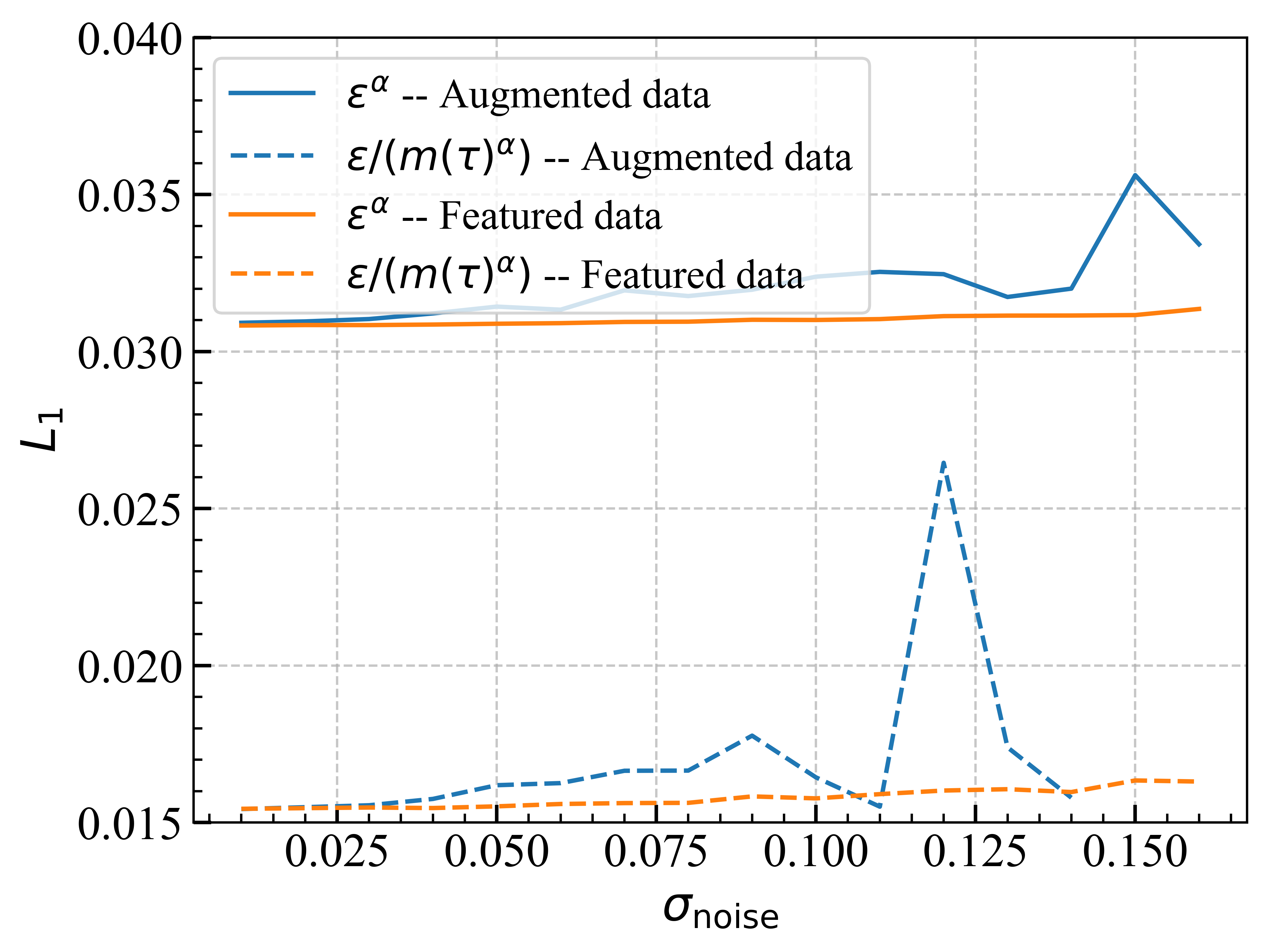}
    \end{minipage}
    \hfill
    \begin{minipage}[b]{0.56\linewidth}
        \includegraphics[width=\linewidth]{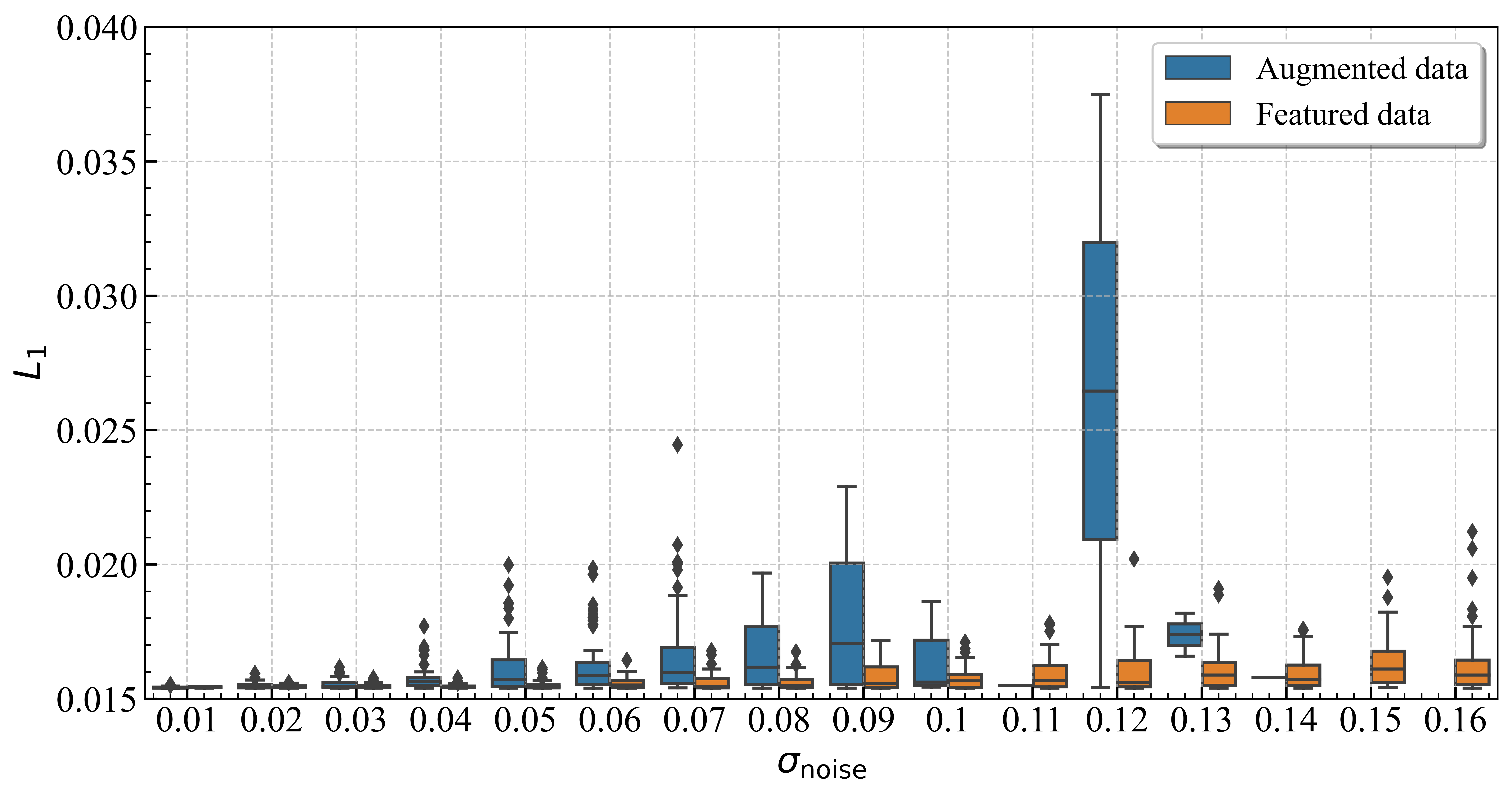}
    \end{minipage}
    \caption{(left) Mean $\Lone$-error and (right) $\Lone$-error distributions for the true ($\mPred=\varepsilon/m(\tau_\text{geo})^{\aPred}$, dashed lines/boxplots) and alternative ($\mPredPrime=\varepsilon^{\aPredPrime}$, solid lines) equations across noise levels, comparing Augmented data (blue) and Featured data (orange).}
    \label{fig:Lone}
\end{figure}

Figure~\ref{fig:Lone} (left) shows the mean $\Lone$-error conditioned on $\mPred$ and $\mPredPrime$, while Figure \ref{fig:Lone} (right) presents boxplots of $\Lone$-error distributions for $\mPred$. 

The results presented in Figure~\ref{fig:FrequencyAndCoeffError} (left) demonstrate the frequency of occurrence of two distinct equations derived using our framework under varying noise levels $\noise$. The true equation type, $\mPred=\varepsilon/m(\tau_\text{geo})^{\aPred}$ and a competing alternative equation, $\mPredPrime=\varepsilon^{\aPredPrime}$ are analyzed across two datasets: "Augmented data" and "Featured data". The key observation is that the true equation type consistently exhibits the highest frequency for the "Featured data" (orange dashed line) across all noise levels. This indicates the robustness of our framework in identifying the true equation, even as noise is increased. The solid orange line, which represents the alternative equation, consistently has a lower frequency, further emphasizing the reliability of the "Featured data" in yielding the true equation.

For the "Augmented data" (blue dashed line), the frequency of the true equation is higher than that of the alternative equation up to a noise level of approximately $\noise$=0.08. Beyond this threshold, the alternative equation (blue solid line) becomes more dominant. This decline in the frequency of the true equation at higher noise levels can be attributed to the method employed in "Augmented data". In this approach, the GAT network generates augmented data by learning the patterns of the original dataset and applying meaningful noise. However, when additional noise is introduced manually, the cumulative effect of both sources of noise may exceed the framework's capacity to accurately identify the true equation. As a result, the SR process tends to favor the alternative equation under these conditions. Figure~\ref{fig:FrequencyAndCoeffError}  (right) illustrates the coefficient error, which measures the deviation of the derived $\alpha$ value from its true value in the equations, as a function of increasing noise levels ($\noise$). The "Augmented" (blue boxes) and "Featured data" (orange boxes) are compared across the same noise levels as in the first plot. 

The "Featured data" (orange boxes) demonstrates consistently lower coefficient error and variance across all noise levels, underscoring its robustness in preserving parameter accuracy. Conversely, the "Augmented data" (blue boxes) exhibits increasing coefficient error and variability as noise levels rise, especially beyond $\noise$=0.08. This trend aligns with the frequency results from Figure~\ref{fig:FrequencyAndCoeffError} (left), where the true equation's frequency decreases at higher noise levels for the "Augmented data". Note that at high noise levels $\noise\geq0.11$ within the augmented data set, the GAT framework is not able to predict the correct equation frequently, resulting in no or no meaningful boxplots.
Overall, the "Featured data" approach outperforms the "Augmented data" in both identifying the true equation and maintaining parameter accuracy across a range of noise levels. While the "Augmented data" is effective at low to moderate noise levels, its performance deteriorates under higher noise due to compounded noise effects. These findings demonstrate the importance of feature engineering and dataset preparation in enhancing the robustness and accuracy of equation discovery within the proposed framework.

The results presented in Figures~\ref{fig:Rsquared} (left) and \ref{fig:Rsquared} (right) further analyze the robustness and accuracy of our framework by focusing on the $\Rtwo$-score of the predicted equations under varying noise levels $\noise$. Specifically, these plots consider two equation types: the true equation $\mPred=\varepsilon/m(\tau_\text{geo})^{\aPred}$ (dashed lines) and the alternative equation $\mPredPrime=\varepsilon^{\aPredPrime}$ (solid lines), across two datasets: "Augmented data" (blue) and "Featured data" (orange).

\smallskip
\noindent

Figure~\ref{fig:Rsquared} (left) illustrates the mean $\Rtwo$-score for these equations as a function of noise. As expected, the $\Rtwo$-score decreases with increasing noise levels for all cases, reflecting the challenge of maintaining accuracy as data becomes noisier. Interestingly, when the true equation $\mPred$ is identified, the "Augmented data" (blue dashed line) consistently achieves higher $\Rtwo$-score than the "Featured data" (orange dashed line). This can be attributed to the meaningful noise added by the GAT framework, which enhances the diversity and generalization of the dataset. Consequently, if the GAT framework identifies the correct equation, it is more likely to accurately predict the $\alpha$ values while minimizing the risk of overfitting. However, as noise levels increase, the compounded noise effects make it harder for the GAT network to frequently identify the true equation, as observed in Figure~\ref{fig:FrequencyAndCoeffError}. On the other hand, the alternative equation $\mPredPrime$ exhibits consistently lower $\Rtwo$-score for both datasets, with the "Featured data" (orange solid line) outperforming the "Augmented data" (blue solid line). This trend aligns with earlier observations, where the "Featured data" approach demonstrates superior robustness and reliability under increasing noise. Figure~\ref{fig:Rsquared} (right) complements this analysis by showing boxplots of the $\Rtwo$-score distribution when the true equation $\mPredPrime$ is predicted. The boxplots highlight that the $\Rtwo$-score for the "Augmented data" (blue) are generally higher and less variable compared to the "Featured data" (orange), especially at lower noise levels. However, as noise increases, the gap narrows, and the "Featured data" maintains better consistency across the noise spectrum. Recall that for noise level $\noise\geq0.11$ the GAT framework is not able to predict the correct equation frequently for the "Augmented data", resulting in no meaningful boxplots.

\smallskip
\noindent
The results in Figure~\ref{fig:Lone} analyze the $\Lone$-error, measuring the deviation between predicted and actual outputs under varying noise levels ($\noise$), and provide further insights into the robustness of the framework. In Figure~\ref{fig:Lone} (left), the mean $\Lone$-error for the true equation $\mPred=\varepsilon/m(\tau_\text{geo})^{\aPred}$ is consistently lower for the "Featured data" (orange dashed line) compared to the "Augmented data" (blue dashed line), indicating superior accuracy and robustness of the "Featured data" approach across all noise levels. The alternative equation $\mPredPrime=\varepsilon^{\aPredPrime}$ exhibits higher $\Lone$-error for both datasets, further confirming its lower reliability. Figure~\ref{fig:Lone} (right) complements this by showing the distribution of $\Lone$-error, where the "Featured data" exhibits smaller variances and consistently lower errors compared to the "Augmented data". As noise increases, the "Augmented data" shows higher $\Lone$-error and wider distributions, consistent with the trends observed in earlier figures. These results reinforce the conclusions from Figures~\ref{fig:FrequencyAndCoeffError}-\ref{fig:Lone}, where the "Featured data" demonstrates greater reliability in identifying the true equation and maintaining accuracy, while the "Augmented data" approach, though occasionally more accurate when it predicts the true equation, is less consistent at higher noise levels.

\noindent
Building on these findings, Figure~\ref{fig:FrequencyAndCoeffError} highlights a trade-off between frequency and accuracy. While the "Featured data" approach excels in identifying the true equation more frequently (Figure~\ref{fig:FrequencyAndCoeffError} (left)) and maintaining lower coefficient error (Figure~\ref{fig:FrequencyAndCoeffError} (right)), the "Augmented data" provides a slightly higher $\Rtwo$-score when the true equation is identified, particularly at lower noise levels (Figure~\ref{fig:Rsquared} (left)). This underscores the complementary strengths of both methods: the "Featured data" for robustness and frequency, and the "Augmented data" for accuracy in parameter prediction. Together, these insights emphasize the importance of balancing noise augmentation and feature engineering to optimize both the discovery and accuracy of SR equations within the framework.

\section{Future extensions }
\begin{figure}[h]
    \centering
   \includegraphics[width=0.7\linewidth]{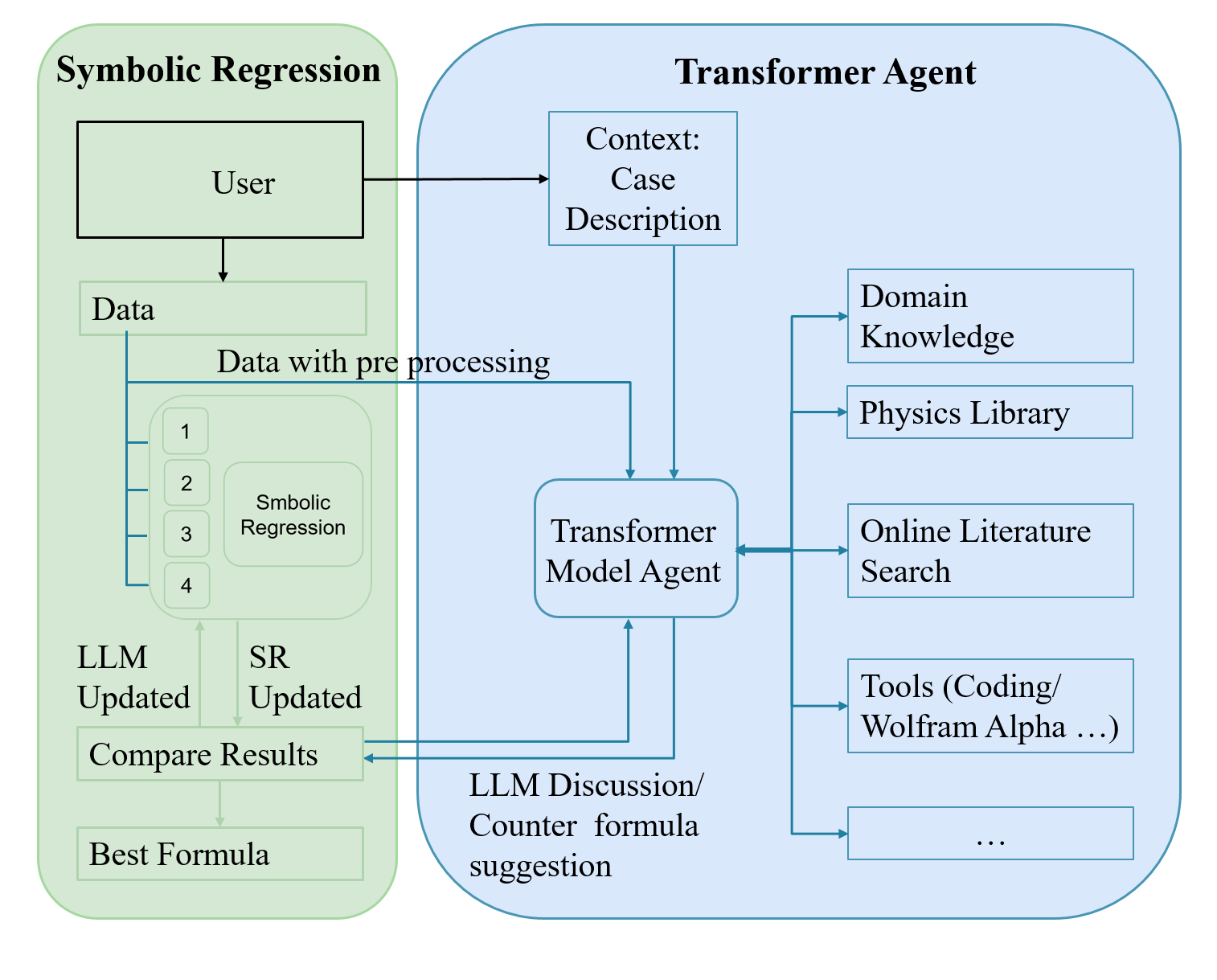}
    \caption{Proposed extension of the framework, attaching iterative mechanistic discussion and modification of SR-based equations in a transformer agent.}
    \label{fig:outlook}
\end{figure}

A future vision or our framework involves the integration of SR with Large Language Models (LLM) or transformer models, specifically the latest generation of so-called "reasoning" models, integrating an internal chain-of-thought process to discuss a matter from different perspectives before generating a final response. This integration enables iterative, domain‐informed discussion and refinement of candidate equations. Transformer models are a novel type of AI capable of contextual interpretation allowing for internal questioning and open‐ended hypothesis formation. These models furthermore can access domain knowledge in a dynamic manner (e.g. via retrieval augmented generation, model fine-tuning, or a mixture of expert approaches), propose modifications to equations, and verify consistency using natural language or code-oriented tools (e.g., Wolfram Alpha or Python libraries). Figure~\ref{fig:outlook} (right panel), shows the envisioned LLM extension. Outputs of the existing SR pipeline, which are candidate equations describing the phenomena to be modeled, are passed to a transformer model agent. This agent can have access to tools such as an internal correlation analysis, real-time references (via online literature searches or local documentation), and a suite of external tools such as mathematics packages. Upon receiving a set of candidate equations, the LLM can inspect key properties, via task-specific promptings, such as dimensional consistency, or boundary conditions specifications. Most importantly, the agent, pre-trained on large corpora of text including physics and other literature and potentially fine-tuned on domain knowledge (e.g. process engineering textbooks) can discuss the equations context-specifically, relying both on its internal training model, external databases for cross-referencing as well as a case definition provided by the user along the data. Based on that discussion counter equations would be provided back to SR. Equations are thus further optimized via oscillation between SR and the transformer agent.

\section{Conclusions}
This study has successfully demonstrated the effectiveness of a hybrid AI framework in improving the prediction accuracy of quantitative structure-property relationships for porous media. The application of this hybrid approach allowed us to derive robust, predictive equations that accurately reflect the complex interplay of microstructural descriptors and their effects on the $M$-factor, which plays a crucial role for materials such as lithium-ion batteries, fuel cells, and fiber-based materials. Our results underscore the ability of the hybrid AI framework to process a huge dataset of 90,000 virtual 3D microstructures, effectively capturing and modeling the essential features that determine material behavior. Through systematic variation of noise levels and the use of rigorous validation techniques, we have ensured that our equations are not only statistically sound, but also practically relevant. Furthermore, the study highlighted the critical role of feature engineering and data enrichment in refining the predictive accuracy of our equations. The use of GATs to improve data quality and feature interaction analysis proved particularly valuable, as it enabled the identification of previously underrepresented but significant microstructural descriptors. Looking ahead, the integration of the hybrid AI framework with emerging technologies such as LLM and transformer-based reasoning systems represents a promising avenue for further improving the applicability of predictive models in materials science. These advances are expected to facilitate the development of next-generation materials with optimized properties tailored for specific applications. 
\section{Acknowledgments }
The authors want to thank the Deutsche Forschungsgemeinschaft (DFG) for funding in the context of the SPP 2364 “Autonomous Processes in Particle Technology” (KI 1839/3, SCHM 997/47-1) as well as the Heisenberg Program SCHI 1265/19-1 “Digital methods for complex systems in process and production engineering” (grant no. 500382045). 

\bibliographystyle{myAbbrvnat}
\bibliography{Sec_7_Bibliography}

\end{document}